\documentclass{iau_FM_tmp}
\usepackage{graphicx}

\newcommand{\lea}{{\>\rlap{\raise2pt\hbox{$<$}}\lower3pt\hbox{$\sim$} \>}}

\title[Angular momentum and galaxy formation] 
{New perspectives on galactic angular momentum, galaxy formation, and the Hubble Sequence}

\author[S.\ Michael Fall \& Aaron J.\ Romanowsky]   
{S.\ Michael Fall$^1$
 \and Aaron J.\ Romanowsky$^2$}

\affiliation{$^1$Space Telescope Science Institute, \\ 
3700 San Martin Drive, Baltimore, MD 21218, USA \\ email: {\tt fall@stsci.edu} \\[\affilskip]
$^2$Dept. of Physics \& Astronomy, San Jos\'e State University, \\ One
Washington Square, San Jose, CA 95192, USA \\email: {\tt aaron.romanowsky@sjsu.edu}}

\pubyear{2018}
\jname{FM6, Galactic Angular Momentum} 
\editors{Danail Obreschkow, ed.}
\begin{document}

\maketitle

\begin{abstract}
This paper provides a summary of our recent work on the scaling relations between the specific angular momentum $j_{\star}$ and mass $M_{\star}$ of the stellar parts of normal galaxies of different bulge fraction $\beta_{\star}$.
We find that the observations are consistent with a simple model based on a linear superposition of disks and bulges that follow separate scaling relations of the form $j_{\star{\rm d}} \propto M_{\star{\rm d}}^{\alpha}$ and $j_{\star{\rm b}} \propto M_{\star{\rm b}}^{\alpha}$ with $\alpha = 0.67 \pm 0.07$ but offset from each other by a factor of $8 \pm 2$ over the mass range $8.9 \leq \log (M_{\star}/M_{\odot}) \leq 11.8$.
This model correctly predicts that galaxies follow a curved 2D surface in the 3D space of $\log j_{\star}$, $\log M_{\star}$, and $\beta_\star$.
\end{abstract}

\firstsection 
\section{Introduction}

Specific angular momentum ($j = J/M$) and mass ($M$) are two of the most basic properties of galaxies.
We have studied the scaling relations between $j$ and $M$ from both observational and theoretical perspectives (Fall 1983; Romanowsky \& Fall 2012; Fall \& Romanowsky 2013, 2018; hereafter Papers 0, 1, 2, and 3).
Here, we present some highlights from Paper 3 of this series.

\section{Results}

Figure~1 shows $\log j_{\star}$ plotted against $\log M_{\star}$ for the 94 galaxies in our sample (with $8.9 \leq \log (M_{\star}/M_{\odot}) \leq 11.8$).
Galaxies of different bulge fraction, $\beta_{\star} \equiv (B/T)_{\star} \equiv M_{{\star}{\rm b}} / ( M_{{\star}{\rm d}} +  M_{{\star}{\rm b}} )$, are indicated by symbols with different shapes and colors in this diagram.
Here and throughout, the subscript $\star$ refers to the stellar components of galaxies, as distinct from their interstellar, circumgalactic, and dark-matter components, while the subscripts d and b refer to disks and bulges, respectively. 
We note from Figure~1 that galaxies with different $\beta_{\star}$ follow roughly parallel scaling relations of the form $j_{\star} \propto M_{\star}^{\alpha}$ with exponents close to $\alpha = 2/3$ ($\alpha \approx 0.6$ for disks, $\alpha \approx 0.8$ for bulges).

\begin{figure}
\begin{center}
 \includegraphics[width=5in]{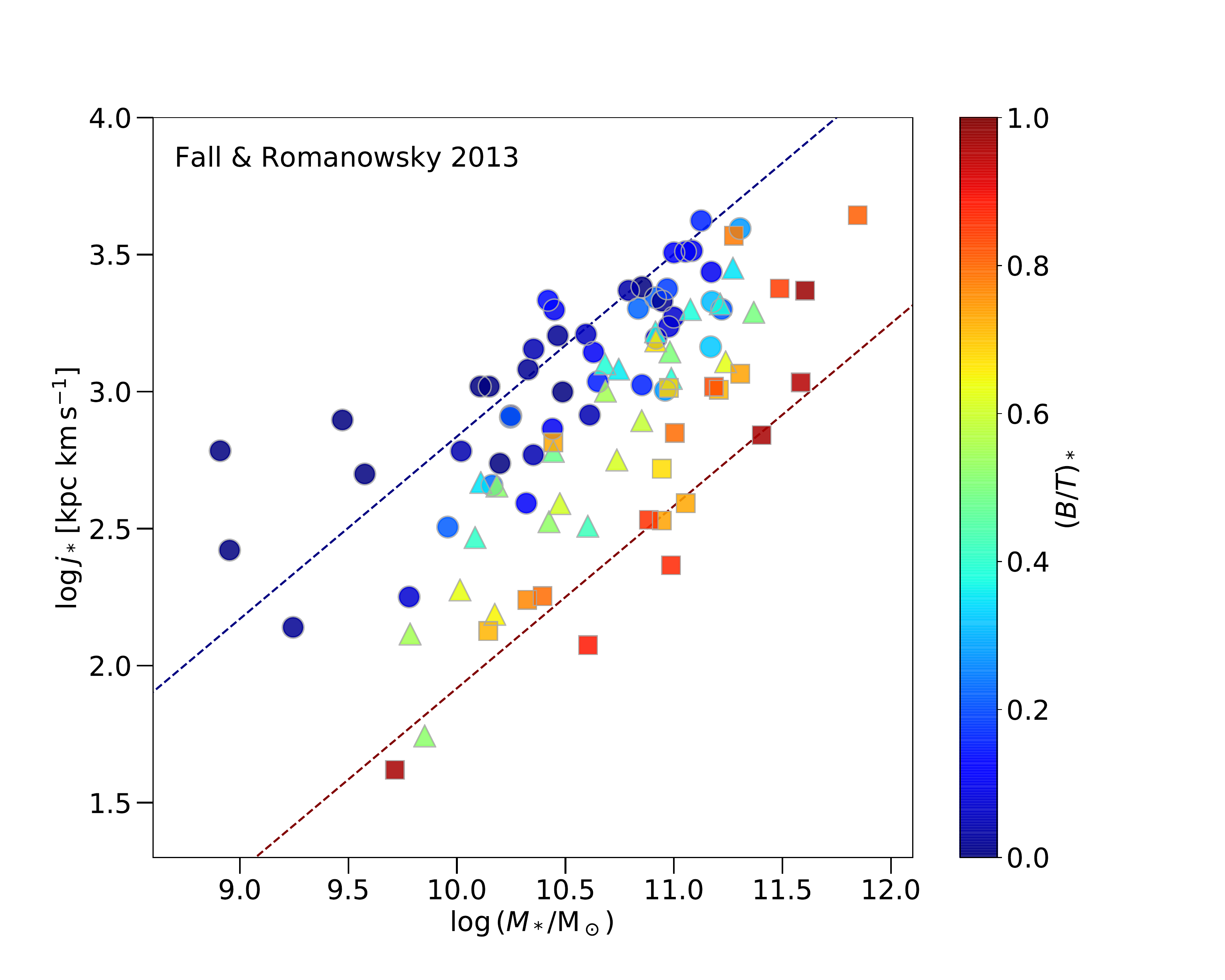} 
 \caption{Stellar specific angular momentum $j_\star$ versus stellar mass $M_\star$ for galaxies of different stellar bulge fraction $\beta_\star \equiv (B/T)_\star$
 (as indicated by symbol shapes and colors).  The dashed lines are scaling relations for disks and bulges from 3D fitting.}
   \label{fig2}
\end{center}
\end{figure}

Figure~2 illustrates schematically the parallel $j_{\star}$--$M_{\star}$ scaling relations for galaxies of different bulge fraction $\beta_*$.
This immediately suggests a connection between the locations of galaxies in the $j_{\star}$--$M_{\star}$ diagram and their morphologies.
And this in turn suggests that the distribution of galaxies of different $\beta_\star$ in the $j_{\star}$--$M_{\star}$ diagram is a physically based alternative to the Hubble sequence.
The analogy here is with the description of elementary particles -- the ``eigenstates'' for galaxies being disks and bulges.
One wonders whether Hubble might have proposed a classification scheme for galaxies based on physical variables like $j$ and $M$ if he had been a physicist rather than an astronomer. 

\begin{figure}
\begin{center}
 \includegraphics[width=5in]{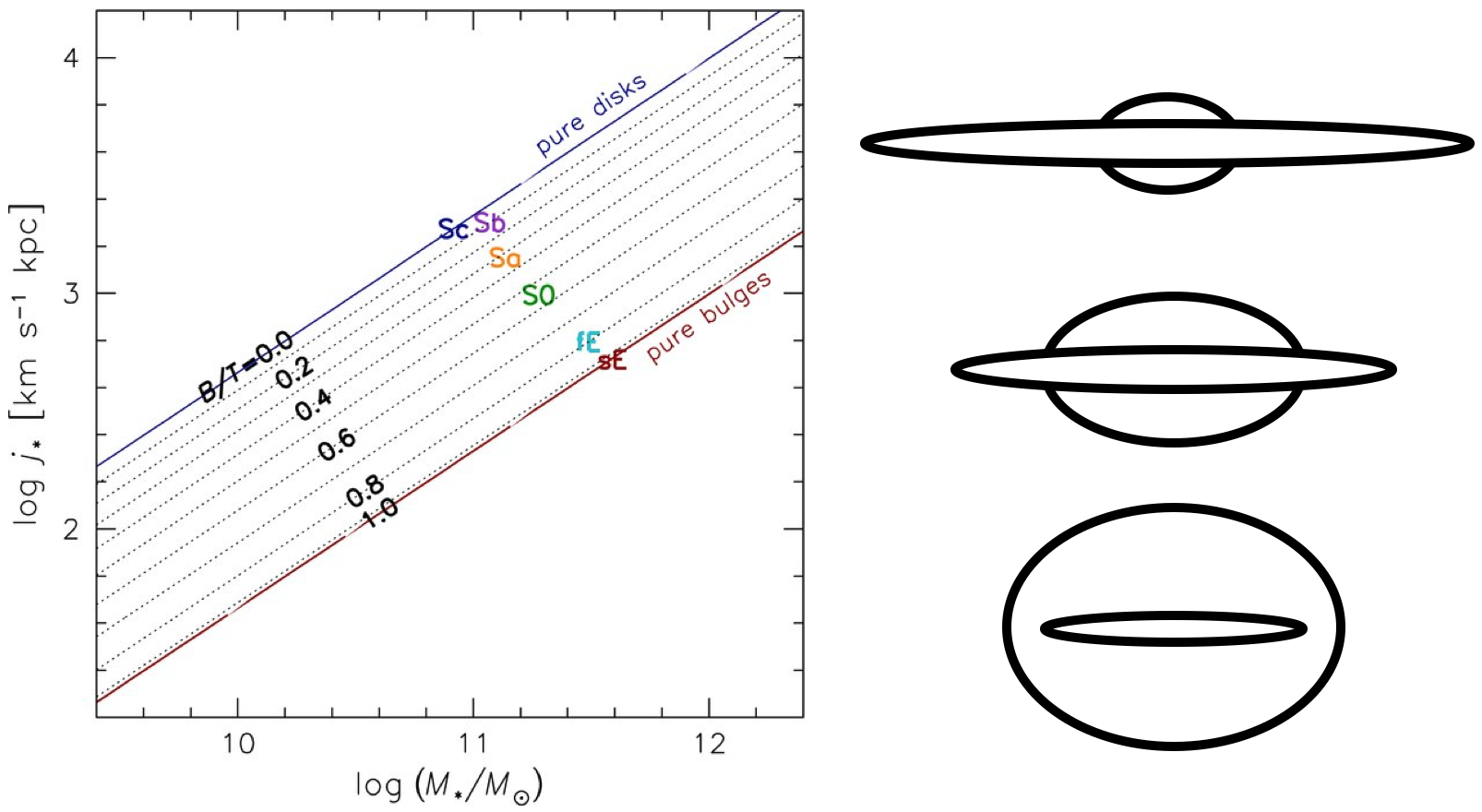} 
 \caption{Physically motivated classification diagram of galaxies, with parallel $j_\star$--$M_\star$ scaling relations
 for fixed bulge fractions (see cartoon examples at right).}
   \label{fig2}
\end{center}
\end{figure}

Figure~3 shows the distribution of our sample galaxies in the 3D space of ($\log j_{\star}$, $\log M_{\star}$, $\beta_{\star}$).
Figure~1 is, of course, just the projection of this distribution onto the $\log j_{\star}$--$\log M_{\star}$ plane.
We note that galaxies lie on or near the curved 2D orange surface in the 3D space. 
The orange surface is derived from a simple model based on a linear superposition of disks and bulges that follow separate scaling relations of the form $j_{\star{\rm d}} \propto M_{\star{\rm d}}^{\alpha}$ and $j_{\star{\rm b}} \propto M_{\star{\rm b}}^{\alpha}$ with $\alpha = 0.67 \pm 0.07$ but offset from each other by a factor of $8 \pm 2$.

\begin{figure}
\begin{center}
 \includegraphics[width=4.5in]{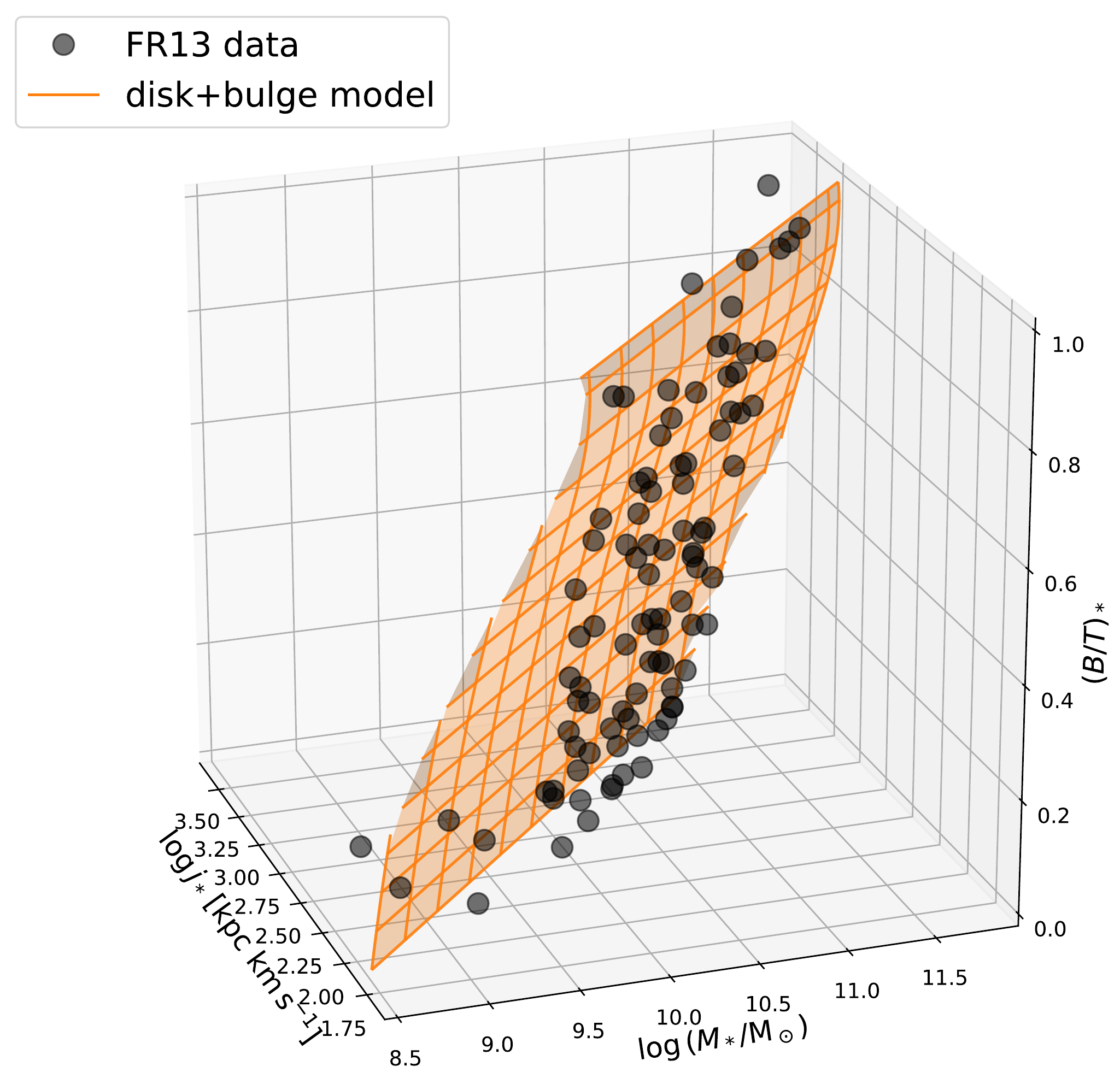} 
 \caption{Bulge fraction versus specific angular momentum and mass. Points show the data, and 
 orange surface shows our 3D relation based on independent disks and bulges.}
   \label{fig3}
\end{center}
\end{figure}
 
In Paper~3, we make detailed comparisons between our $j_{\star}$--$M_{\star}$ scaling relations and those of other authors. 
We find excellent agreement between our results from Paper~2 and those of Obreschkow \& Glazebrook (2014) and Posti et al.\ (2018) for disk-dominated galaxies.
The $j_{\star}$--$M_{\star}$ scaling relation derived by Sweet et al.\ (2018) appears to suffer from an unknown systematic error (by a factor of 2) relative to the relations derived in the other three studies.
We find no statistically significant indication that galaxies with classical bulges and pseudo bulges follow different relations in ($\log j_{\star}$, $\log M_{\star}$, $\beta_{\star}$) space.

In Paper~3, we provide an updated interpretation of the $j_{\star}$--$M_{\star}$ scaling relations, following the precepts of Paper~1.
In particular, we have revised slightly our earlier estimates of the fractions of angular momentum in the stellar components of galaxies relative to dark matter, $f_j \equiv j_\star / j_{\rm halo}$.
We now find $f_j \sim 1.0$ for disks (slightly higher than before) and $f_j \sim 0.1$ for bulges (slightly lower than before).
We also note that these fractions are expected to be nearly constant over the mass range $10^{9.5} M_\odot \lea M_\star \lea 10^{11.5} M_\odot$.
Posti et al.\ (2018) suggested that $f_j$ may decrease gradually toward lower galactic masses based on their extension of the $j_{\star}$--$M_{\star}$ relation down to $\sim10^7 M_\odot$.
Future studies should aim to determine $f_j$ for dwarf galaxies from the {\it baryonic} $j$--$M$ relation (including both stars and cold gas) since this may be slightly shallower than the {\it stellar} $j$--$M$ relation, and thus consistent with $f_j \approx {\rm constant}$. 

We note that the retention factor $f_j \sim 1.0$ derived from the observed $j_{\star}$--$M_{\star}$ relation for galactic disks agrees well with the value of $f_j$ postulated in simple disk formation models (Paper~0), although the physical reasons for this agreement are still an active research topic (as discussed at this meeting by Bullock, DeFelippis, El-Badry, Genel, and others).    
The retention factor $f_j \sim 1.0$ also agrees well with the observed sizes of disk-dominated galaxies over the redshift range $0 \leq z \leq 3$.
Using the method of abundance matching, Huang et al.\ (2017) showed that the relation between the sizes of galaxies and their dark-matter halos is linear and stable over this redshift range and consistent with simple disk formation models (i.e., $f_j \sim 1.0$).

\section{Conclusions}

1.  The observed $j_{\star}$--$M_{\star}$ scaling relations for galaxies with different $\beta_\star$ constitute a physically motivated alternative to subjective classifications schemes such as the Hubble sequence.

2.  At fixed $\beta_\star$, specific angular momentum and mass are related by power laws, $j_\star \propto M_{\star}^{\alpha}$, with $\alpha \approx 0.6$ for disks, $\alpha \approx 0.8$ for bulges, and $\alpha \approx 2/3$ overall. 

3.  For giant galaxies (with $10^{9.5} M_\odot \lea M_\star \lea 10^{11.5} M_\odot$), the angular momentum retention or sampling factors are $f_j \sim 1.0$ for disks and $f_j \sim 0.1$ for bulges.

\end{document}